# Telescope time without tears – a distributed approach to peer review

**Michael Merrifield and Donald Saari describe a new procedure designed to take the stress out of allocating the limited time available on telescopes.**

Abstract
The procedure that is currently employed to allocate time on telescopes is horribly onerous on those unfortunate astronomers who serve on the committees that administer the process, and is in danger of complete collapse as the number of applications steadily increases. Here, an alternative is presented, whereby the task is distributed around the astronomical community, with a suitable mechanism design established to steer the outcome toward awarding this precious resource to those projects where there is a consensus across the community that the science is most exciting and innovative.

Most of the cutting-edge telescopes that are in use around the World today are heavily oversubscribed, with typically five or ten times more nights needed to carry out the projects that astronomers want to do than are available. This valuable limited resource is therefore usually allocated by a rather complicated competitive process in which individuals or research collaborations submit applications describing the projects they wish to carry out; these applications are then assessed by a review panel of maybe five or six astronomers, and the best of them are approved and allocated time on the telescope.

One of the authors (MRM) has been motivated to contemplate the shortcomings of this procedure by the presence on his desk of a bulging file of 113 telescope applications that he had been sent to assess as a member of the European Southern Observatory (ESO) Observing Programme Committee. In a desperate attempt to put off burying himself under this mountain of paperwork, he got to thinking about the way we assess telescope applications, and was struck by a number of points about the current process.

First, it is not reasonable to place this burden of assessment on such a small fraction of the community. Although the load averages out over time to an extent as members cycle on and off panels, some people repeatedly duck their share of the work, or do such a bad job of it the first time that they are never invited again, so the work is not fairly distributed even over the long term.

Second, no matter how good one's intentions, one simply cannot do as good a job as the task deserves. ESO allows three weeks to complete the task of assessing these 113 proposals, on top of the typical university lecturer's full schedule of teaching and the occasional grabbed moment for research. Even paring things down to 15 minutes assessing each proposal, and five minutes to assign a score and write up some comments, we are still talking about an entire working week doing nothing else, on top of the several days of meetings required to thrash out a finalized list through panel meetings. And even if we could compromise on the work that we are actually being paid to do sufficiently to create this much time, it would still be impossible to do a satisfactory job, as there is no way that anyone can maintain a big picture sense of the quality of such a large pile of submitted proposals: it is impossible to control against drifts in the grades assigned to allow a fair comparison between Proposal 1 and Proposal 113, which is required for them to be ranked properly against each other.

Third, there is really very little control to ensure that those involved take the process seriously. Indeed, as mentioned above, there is actually a disincentive to do so, since acquiring a reputation for doing this job well simply invites more of it to be heaped upon you. There is obviously some benefit to the individual who serves on such panels, since one gets an interesting overview of what is currently exercising the broader astronomical community, obtains valuable insights into the peculiarities of the organization's particular system, and also learns some of the basic pitfalls into which a surprisingly large number of applications fall (how can anyone write an application to observe 317 stars without somewhere demonstrating that one could not achieve the main science goals with 12 objects?). However, it would be hard to argue that these returns outweigh the costs in terms of time, and in fact the benefits accrue no matter how bad a job one does of the reviewing. Ultimately, the system relies to an uncomfortable extent upon panel members' work ethics, and the rather crude incentive of not wanting to appear ill-informed in front of their peers on the panel.

Fourth, it should be borne in mind that there is no objective right answer in this kind of peer review process. Ideally, one would like to produce a single ordered list that ranks the scientific merit of the individual proposals, so that one can work down the list assigning the available telescope time to the best applications. However, this ideal is at best an over-simplification and at worst a complete fantasy. Even within a single sub-field of astronomy, there is no single, objective way to weigh up a highly speculative proposal that would revolutionize the subject if it paid off against a more run-of-the-mill incremental study. When one then tries to compare disparate topics, there is absolutely no rational basis for deciding whether a definitive study of, say, quasar activity is more or less important than one on exoplanets. The unachievability of an objective ranking of applications does not mean that one should not try, but it does imply that it is not worth investing ever greater efforts to improve the precision of the ranked list when there is such a fundamental limitation on its accuracy.

Finally, this issue is not going to go away. For the reasons outlined above, we would argue that the system is already broken in that it is unreasonably burdensome and less than optimal in its outcome because of its excessive load, and this situation is only likely to get worse with time: ESO's total applications have gone up from around 500 per half-year a decade ago to more than a thousand in the latest round, and there are no signs that this number is saturating. Indeed, there is a real danger of an element of positive feedback, in that as the system becomes more overloaded it becomes less effective, so applicants recognize that there is a larger random element in winning telescope time, and accordingly "buy more lottery tickets" by putting in larger numbers of applications.

So, how does one go about improving on the current overly-burdensome system? Clearly, something rather radical is required to stop us from sinking under an ever-expanding sea of assessment, but any new system should still meet a number of basic principles:

- Some incentive should be put in place to reduce the pressure toward ever more applications.
- The workload of assessment should be shared equitably around the community.
- The burden on each individual should be maintained at a reasonable level so that it is physically possible to do a fair and thorough job.
- There should be some positive incentive to do the task well.
- The ultimate ranked list of applications should be as objective a representation of the community's perception of their relative scientific merits as possible.

Many of the issues involved are clearly not specific to astronomy, and there already exists a great deal of literature on "mechanism design" to optimize this kind of process. In his attempts to put off assessing telescope proposals by learning more about the subject, MRM came into contact with this work's co-author (DGS) who has extensive expertise in both astrophysics and the mathematical theory of voting and its associated complexities. The remainder of this paper is a proposal arising from the discussions that the authors have held, which is designed to be less onerous on any individual, but still meet the above principles, and, in particular, result in a ranked list of telescope applications that as closely as possible reflects the community's view of their relative merits.

## A distributed approach to peer review

Here, then, is an outline proposal for a process to replace the current peer review model for telescope time allocations:

1. As now, principal investigators (PIs) submit proposals against a set deadline, specifying in which sub-field their applications lie.
2. After the deadline, each of these $N$ PIs is sent $m$ proposals from other applicants in their sub-field.
3. If the PI has a conflict of interest (institutional or personal) on any application, he declares it as in most current refereeing processes, and the application is replaced by an alternate.
4. The PI assesses these $m$ applications, and compiles a ranked list, placing them in the order in which she thinks the community would rank them, *not* her personal preferences.  How she carries out this process is up to the PI: she could, for example, call on the combined views of her co-applicants, or delegate the task to one co-investigator.  As now, she is not allowed to communicate with the applicants on the proposals she is assessing.
5. The PIs all submit their ranked list of $m$ applications.  Failure to carry out this refereeing duty by a set deadline would lead to the PI's own proposal being automatically removed from the ranking: the refereeing element should be viewed as much a part of the application as any other, and not carrying it out means that the proposal is incomplete and should be rejected.
6. These individual sub-lists of rankings are then combined to produce an optimized global list ranking all $N$ applications.
7. Finally, each PI's individual rankings are compared to the positions of the same $m$ applications in the globally-optimized list.  If both lists appear in approximately the same order, then the PI has done a good job and is rewarded by having his application moved a modest number of places up the final list relative to those from PIs who have not predicted the community's view so well.

At this point, the lists from different sub-fields could be handed over to a conventional telescope allocation committee to merge the lists with some degree of human intervention, as the amount of work involved is now reduced to very manageable proportions: the allocation panel has simply to work down their ranked lists, merging them appropriately, so the work involved scales only with the available telescope time, not the number of applications.  However, one could also imagine simply going down each list allocating pro-rata on the historic record of the number of applications or hours requested in each sub-field, which, in practice, is what tends to happen anyway.

This process seems to at least meet the general principles set out above:

- It shares the workload widely around the community, weighting appropriately towards those who submit the largest number of applications.
- This weighting also provides a disincentive against the lottery-ticket approach to telescope applications, or at the very least makes the individual applicants take responsibility for the extra workload that this approach generates.
- The load on any individual does not go up if the number of applications increases.
- If the value of $m$ is chosen appropriately and the combining algorithm is designed robustly, the system has sufficient redundancy to be insensitive to outliers and attempts to subvert the process.

- It explicitly seeks to produce a list that reflects the views of the entire community, both by spreading the process more widely and by focusing referees' minds on considering the broad interests of the community rather than their individual preferences.
- It provides a transparent, positive incentive to do a good job.

However, as usual, the devil is in the detail. Below, we delve a little deeper into the practicalities of implementing such a scheme, and present some simulations to test whether it might work in practice.

## Creating the global ordered list

The first challenge in this process involves the combination of the individual PIs' lists into the globally-optimized rank order. In effect, this is an electoral process, and there is a long and extensive literature on what the optimum method is for deciding the outcome of such an election from the individual votes, which we can draw on for this application.

Interestingly, there is already an astronomical connection with this literature, in that one of the seminal contributions was made in the late eighteenth century by Jean-Charles de Borda. As well as being involved in the definition of the metre based on estimates of the Earth's circumference and designing various related astrometric instruments (and, incidentally, recalculating all the trigonometric tables based on a circle divided into 400 degrees), Borda also came up with an algorithm for combining voters' individual preference lists to decide who wins an election (Borda 1781). In a slate of $N$ candidates, this "Borda count" involves assigning $N-1$ points for each first preference vote cast, $N-2$ for each second choice, down to 0 for a last choice. This scoring mechanism in fact just comprises comparing the candidates in pairs and counting up the number of contests each candidate wins. Suppose, for example, there were three candidates, A, B and C, which a voter chose to rank in this order. If the voter mentally carried out pair-wise comparisons between all possible combinations, A would win two contests (beating both B and C), B would win one contest against C, and C would not score any wins at all. One then simply sums the Borda counts assigned to each candidate by every voter, with the winner of the election being the candidate with the highest score, and so on down the list of runners up to create a global ranked list. In the case of the current process, where each assessor only ranks $m$ applications, one adopts a "modified Borda count" (MBC) in which the points allocated count down from $m-1$ to 0 in each sub-list, giving each application a total score between 0 and $m(m-1)$, which can be mapped into the range $0 < MBC < 1$ by dividing by $m(m-1)$, ordered and ranked.

For the purpose of allocating telescope time, the Borda count has the great benefit that it reflects the broad consensus of the community rather than being driven by simple majority voting. It also has the strength that it is intrinsically relatively robust against attempts to distort the results by tactical voting. However, it should be borne in mind that there is also a result known as the Gibbard–Satterthwaite Theorem, which proves that any fair voting system in which there are more than two possible outcomes can always be subverted by tactical voting (Gibbard 1973, Satterthwaite 1975). Although astronomers are largely an honest bunch of individuals who would not sink to such underhand tactics, one would like to do whatever is possible to remove even the possibility of such a subversion of the allocation process, which is where the concept of rewarding good refereeing comes in.

## Rewarding good refereeing

Once again, there is already a body of research into designing incentives into a process to encourage the participants to behave honestly – a topic known generally in economics and game theory as "mechanism design." Indeed, some of the early work done in this field was specifically directed toward mitigating the effects of the Gibbard–Satterthwaite Theorem by introducing a system of punishments and rewards in voting systems (Groves & Ledyard 1977). In the current context of time allocation on telescopes, where the voters are also the stakeholders, one can design particular effective incentives into the process: in a traditional liberal democracy, the worst that one can ethically do to voters who try to "cheat" through tactical voting is to arrange for their votes to carry less weight in the final tally, but here we can design a system that actively encourages the voters to consider the broader "good of society" through the incentive of increasing the chances of obtaining the direct personal gain of telescope time if they do so.

This idea of rewarding desirable behaviour is not unprecedented in the allocation of telescope time at many observatories. When astronomers build a new instrument to mount on a telescope, the benefit that they bring to the broader community by making the instrument generally available is frequently rewarded with what is known as "guaranteed time," whereby they are "paid" with an allocation of observing time on the telescope. Thus, the principle is already established that observatories reward socially desirable behaviour and a job well done with time on the telescope, so why shouldn't referees also receive a modest boost to their observational programmes for doing a good job in the assessment process?

It is also worth noting that minor reordering of the final list of applications through the distribution of relatively small rewards has very little impact on the global validity of the final list: proposals near the top of the list will be approved irrespective of such modest rearrangement, and those near the bottom will remain unsupported; it is only those in a "penumbra" of intermediate ranks that may change state on the basis of such minor incentives, and the general view is that the process is intrinsically fairly arbitrary when it comes to deciding which such mid-ranking applications should be supported.

There are many ways in which one can assess how well each individual has carried out her task by predicting the final outcome in the global ordered list, which can be tailored to what is deemed important. If, for example, a telescope is oversubscribed such that only a fraction $f$ of applications can be awarded time, the ordering of the bottom of the list real does not matter very much, so one might reward an assessor for placing the same $f \times m$ applications at the top of his list that appear at the top of his sub-set of applications in the global list. One particularly simple figure of refereeing merit, $Q_i$, for applicant $i$ could be derived from the absolute deviation of the rank order of his sub-list from the relative rank order in which the same applications appear in the global list,

$$Q_i = 1 - \frac{1}{\text{int}\left(\frac{1}{2}m^2\right)} \sum_{\substack{\text{applications} \\ \text{in } i\text{'s list} \\ j=1}}^{m} \left| \text{Rank of } j \text{ in } i\text{'s sub-list} - \text{Rank of } j \text{ amongst these } m \text{ in global list} \right|,$$

so if $i$ is a "perfect" assessor who exactly reflects the views of the community he would score $Q_i=1$, while the normalization has been chosen such that if he is an "evil" assessor who deliberately placed the applications in exactly the wrong order then he would score $Q_i=0$.

One further nuance that could be implemented with a modest increase in computation would be to recalculate the global list excluding applicant $i$ in each calculation of $Q_i$. Such a modification would effectively remove any influence of the individual on the averaged consensus when comparing the two to assess the quality of refereeing. One could also envisage iterating the process, whereby a new ranked list is created with any referees with very low values of $Q_i$ excluded, thus producing an ordering that depends only on competent conscientious referees, more closely representing the consensus view of the community.

Having calculated such a figure of merit, one has to decide what to do with it. Since the purpose of this exercise is to reward those who take their refereeing duties seriously, but also potentially to penalize those who try to manipulate the process, one simple algorithm would be to move those who score above some critical value, $Q_{good}$, $n$ places up the list, and to move all those who score less than some other critical values, $Q_{bad}$, $n$ places down the list, relative to those who score somewhere in the middle. The size of $n$ would need some calibration, since one needs it to be large enough for applicants to view it as sufficient carrot to take it seriously, but not so large that it ends up dominating over the relative scientific merits of applications. One natural choice would be to relate it to the scatter amongst the rankings of different assessors, since moves on this scale would be within the intrinsic uncertainty in the process, and so are justifiable as not unduly distorting the peer review process.

## A simulated allocation process

As a test of whether this process might work in practice, we have set up a simulated telescope application round implementing the above procedure. We have assumed that there are $N=100$ applicants, approximately in line with the number of applications that might be assessed within a single panel at present. The true ordering of these applications by merit is specified in the simulation, and we model the intrinsic uncertainty in each referee's personal assessment of these merits by assuming that she is fundamentally unable to distinguish between applications within $\Delta n = 10$ places of each other in this list, but otherwise can rank each application fairly against its competitors.

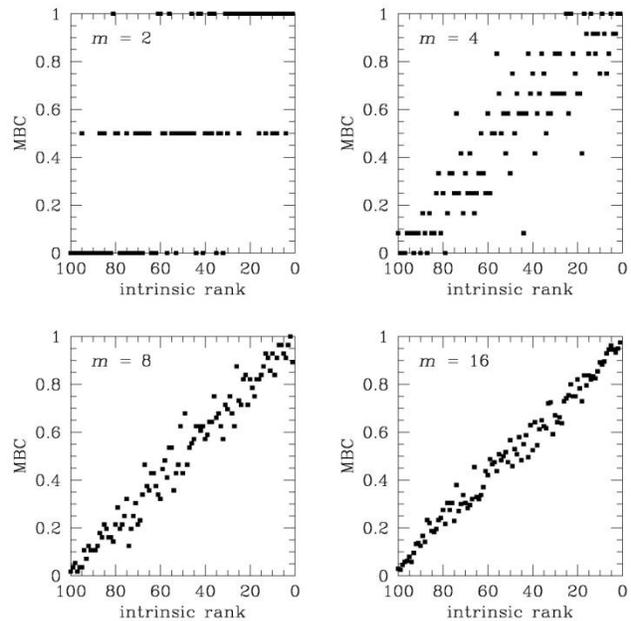

1: Modified Borda Count returned when a sample of 100 applications of known rank are sent to differing numbers $m$ of the applicants for assessment.

Figure 1 shows what happens as one varies the number of applications $m$ that each assessor is asked to rank. Surprisingly, even if each participant only receives two applications to assess, the Modified Borda Count does a reasonable job of combining all the assessments and picking out the top applications, although clearly with significant quantization issues. As $m$ increases, the scatter comes down so that by $m=8$ the error that the process might make is comparable to the intrinsic uncertainty. It is also interesting to note that the scatter is smaller at the extremes of the distribution, as the intrinsic uncertainty in ranking becomes a one-sided process, so there are fewer possibilities for mis-ranking. This effect is particularly advantageous in heavily-oversubscribed application processes, as almost all of the successful applicants are likely to be drawn from the tightly-correlated upper tail.

These simulations assume the ideal case where all assessors are honest and do a perfect job to within the intrinsic ten-place uncertainty. Figure 2 shows what happens in a somewhat more realistic scenario, where 85% of referees are assumed to have this level of competence, 10% are so incompetent that they assign random ranks, and 5% are "evil," and deliberately reverse the order of their rankings; while still somewhat idealized, such a distribution gives a sense of what impact these failing will have on the system. Figure 2 also shows the improvement that one obtains in the final rankings by clipping out the less competent referees from the modified Borda count on the basis of the $Q$ values obtained by comparing the individuals' rankings with the global list.

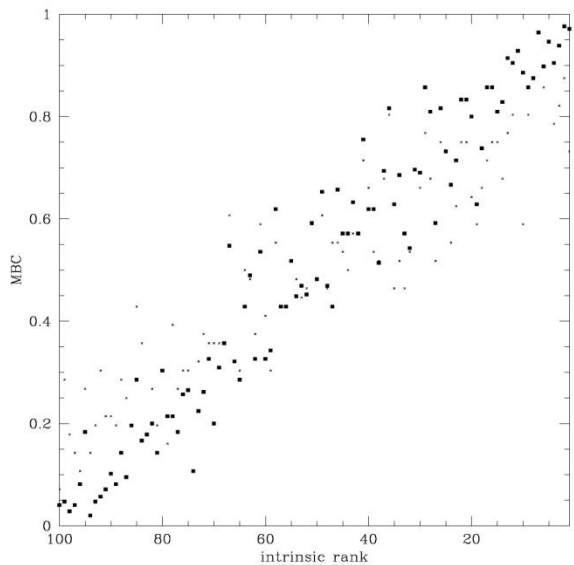

2: Modified Borda count returned in a rather more realistic mix of 85% competent, 10% incompetent, and 5% deliberately mis-ranking referees. The solid points show the improvement in fit obtained after the identified "evil" referees are clipped from the scoring.

Figure 3 shows the associated distribution of values of $Q_i$, demonstrating that it is fairly straightforward to use such a measure to distinguish between malicious, incompetent and competent referees. The separations of these classes also provide an indicator as to the critical values of $Q$ that one might consider adopting to encourage applicants to take the refereeing process seriously: one could, for example, move all applicants that score $Q > 0.6$ up the list, and all those that score $Q < 0.3$ down the list, relative to those in the mid-range. These plots even provide some indication as to how far it is appropriate to move applications in the list: since a move of up to ten places up or down the ranking is within the intrinsic scatter in the process, a move of up to this magnitude could provide a significant incentive without unduly distorting the outcome. Note also that such readjustments will have very little impact on the final list since the vast majority of referees have done an honest job so will not shift in rank relative to each other, but since the basic aim of this incentive is to encourage applicants to take the refereeing seriously, such a null impact is a desirable outcome in itself.

## Limitations

Although this approach has benefits in terms of the manner in which the workload is distributed and the way it seeks to define a community consensus, it also has its shortcomings. These shortcomings need to be addressed, and mechanisms considered to ameliorate them.

For a start, this process is to some extent still dependent on the good will and honesty of the participants. Since there is no direct scrutiny of each referee's actions and no requirement to justify a ranking, there may be a temptation to not take the process seriously or even try to subvert it. However, there is still the same obligation to declare conflicts of interest and otherwise behave in an ethical manner. Further, the system more actively discourages cheating since, for example, anyone deliberately mis-grading a competitor's application would end up damaging their own chances when their grades were found to be out of line with those of their peers.

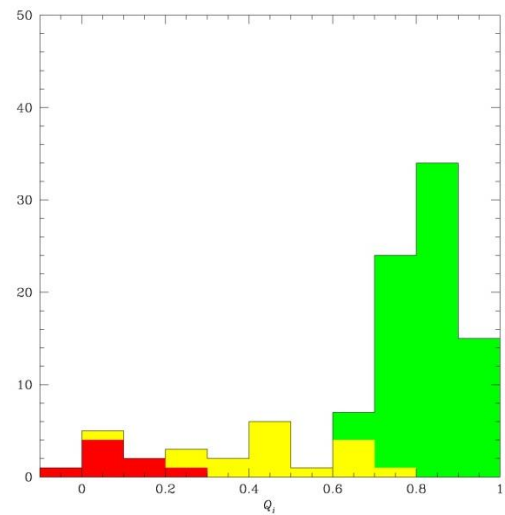

3: Distribution of measured refereeing quality as defined by the $Q$ parameter. Intrinsically competent referees are coloured green, incompetent ones are yellow, and evil ones are red.

There is also the ethical matter of not stealing ideas from other peoples' proposals that one is sent to assess, but this issue is really no different in a distributed model than it is in any other system of refereeing.

One aspect of the current process that this method cannot reproduce is in providing feedback on individual proposals, beyond recording where in the final ranked list they lie. However, the feedback provided on the huge list of applications that most allocation committees assess is necessarily terse and often not very helpful.  Its loss would be offset by a process whereby applicants look at a significant range of other peoples' proposals in every round, allowing them to see how others make their cases convincing.  A related concern is how more junior applicants such as students will be able to handle the refereeing requirement with little past experience.  However, since the purpose here is to integrate the refereeing into the application process, there is a natural mechanism for mentoring such applicants: just as more senior applicants on a proposal would be expected to provide input into writing a convincing case, so they would be expected to help train their more junior colleague in the assessment part of the process.

There is also no mechanism for the referees to present "killer arguments" such as pointing out that the science was done back in 1982 or that the object is in the wrong hemisphere.  More technical aspects such as the object's latitude can be dealt with in an automated fashion, but less obvious show-stoppers are harder to deal with.  However, experience on conventional allocation committees would suggest that such issues are rare and seldom clear-cut, and usually depend on the strength of personality of individuals on the panel as to whether or not the case is made.  There is something rather disturbing about relying on the advocacy skills of individuals, and indeed on whether or not there happens by chance to be a person with the right specialized knowledge on such a small panel, to make such decisions, so it is not clear that a more distributed system necessarily produces an inferior solution.  One could, in any case, envisage a final stage of the process in which the approved proposals are sent for vetting by expert referees with the sole purpose of confirming that such impediments do not exist: since only a small fraction of the total number of applications would be subject to this final hurdle of expert scrutiny, the workload involved would be very manageable.

Perhaps the greatest potential concern is that this procedure will drive the allocation process toward conservative mediocrity, since referees will be dissuaded from supporting a long-shot application if they think it will put them out of line with the consensus.  This effect would need careful monitoring, but it could also be addressed explicitly in the instructions given to applicants as to the criteria they should apply in their assessments: if they are aware that all applicants have been encouraged to support scientifically-exciting speculative proposals, then they are likely to rank such applications highly themselves to remain in line with the consensus.

## Conclusions

In this paper, we have presented a new distributed mechanism designed for assessing applications for observing time on major telescopes, but the method could also be used in many other academic contexts, such as the distribution of grant resources, or even deciding how to allocate limited talk slots in an over-subscribed conference so as to build a programme that most interests the participants.  It represents a rather radical departure from the current centralized mechanism for allocating telescope time, but it is clear that the present system is far from perfect and is painfully burdensome.  Given the inherently subjective nature of peer review, there really is no objective way to claim that this approach would be any better or worse than what we have at the moment in terms of outcome, while it is demonstrably more equitable and less onerous as a process.  However, since no-one has yet taken the bold step of implementing such a procedure, one of the authors cannot put off his return to poring over that fat file any longer…

*Michael R. Merrifield, School of Physics & Astronomy, University of Nottingham, NG7 2RD, UK.*
*Donald G. Saari, Institute of Mathematical Behavioral Sciences, University of California, Irvine, CA 92697, USA.*